\documentclass[sigconf]{acmart}
\AtBeginDocument{%
  }

\setcopyright{acmlicensed}
\copyrightyear{2018}
\acmYear{2018}
\acmDOI{XXXXXXX.XXXXXXX}
\acmConference[Conference acronym 'XX]{Make sure to enter the correct
  conference title from your rights confirmation email}{June 03--05,
  2018}{Woodstock, NY}
\acmISBN{978-1-4503-XXXX-X/18/06}

\usepackage{amsmath,amsfonts}
\usepackage{algorithmic}
\usepackage{graphicx}
\usepackage{textcomp}
\usepackage{xcolor}
\usepackage{algorithm}
\usepackage{subfigure}
\usepackage{diagbox}
\usepackage{threeparttable}
\usepackage{multirow}
\usepackage{color}
\usepackage{bbold}
\usepackage{graphicx}
\usepackage{makecell}
\usepackage{caption}
\usepackage{boldline}
\usepackage{colortbl}
\usepackage{wrapfig}
\usepackage{hyperref}
\usepackage{adjustbox}

\begin{document}

\title[\resizebox{4.5in}{!}{Towards Fair Machine Learning Software: Understanding and Addressing Model Bias Through Counterfactual Thinking}]{Towards Fair Machine Learning Software: Understanding and Addressing Model Bias Through Counterfactual Thinking}

\author{Zichong Wang}
\email{zwang114@fiu.edu}
\affiliation{%
	\institution{Florida International University}
	\city{Miami}
	\state{Florida}
	\country{USA}
	\postcode{33199}
}

\author{Yang Zhou}
\email{zyang@smu.edu.sg}
\affiliation{%
	\institution{Singapore Management University}
	\city{Singapore}
	\country{Singapore}
	\postcode{188065}
}





\author{David Lo}
\email{davidlo@smu.edu.sg}
\affiliation{%
	\institution{Singapore Management University}
	\city{Singapore}
	\country{Singapore}
	\postcode{188065}
}

\author{Wenbin Zhang}
\email{wenbin.zhang@fiu.edu}
\affiliation{%
	\institution{Florida International University}
	\city{Miami}
	\state{Florida}
	\country{USA}
	\postcode{33199}
}

\renewcommand{\shortauthors}{Trovato et al.}

\begin{abstract}
  A clear and well-documented \LaTeX\ document is presented as an
  article formatted for publication by ACM in a conference proceedings
  or journal publication. Based on the ``acmart'' document class, this
  article presents and explains many of the common variations, as well
  as many of the formatting elements an author may use in the
  preparation of the documentation of their work.
\end{abstract}




\renewcommand{\shortauthors}{Zichong Wang et al.}

\begin{abstract}
The increasing use of Machine Learning (ML) software can lead to unfair and unethical decisions, thus fairness bugs in software are becoming a growing concern. Addressing these fairness bugs often involves sacrificing ML performance, such as accuracy. To address this issue, we present a novel counterfactual approach that uses counterfactual thinking to tackle the root causes of bias in ML software. In addition, our approach combines models optimized for both performance and fairness, resulting in an optimal solution in both aspects. We conducted a thorough evaluation of our approach on 10 benchmark tasks using a combination of 5 performance metrics, 3 fairness metrics, and 15 measurement scenarios, all applied to 8 real-world datasets. The conducted extensive evaluations show that the proposed method significantly improves the fairness of ML software while maintaining competitive performance, outperforming state-of-the-art solutions in 84.6\% of overall cases based on a recent benchmarking tool.

\end{abstract}
\maketitle

\section{Introduction}
\label{sec:intro}

Machine learning (ML) systems have garnered widespread recognition due to their demonstrated ability to tackle a wide range of critical tasks. Such tasks include human resource management~\cite{chan2018hiring}, healthcare~\cite{rasmy2021med}, sentiment analysis~\cite{hoang-etal-2019-aspect}, and more. 
The success of ML systems is largely due to the availability of large-scale datasets, as highlighted by a report by Worldbank, which indicates that credit bureaus have been using machine learning for credit scoring and fraud detection purposes. 
Furthermore, ML systems are also employed in the realm of hiring, including resume screening and candidate assessment.
However, despite the positive aspects of ML systems, recent studies have revealed a significant drawback - \textit{ML systems can be biased}. Specifically, ML systems have been shown to exhibit biases in various domains such as gender and race. For instance, state-of-the-art sentiment analysis tools often predict texts containing female names as negative~\cite{biasfinder}. Additionally, a deployed ML system that was used to evaluate a defendant's risk of future crime was found to be biased against black defendants (i.e., producing high risk)~\cite{angwin_larson_kirchner_mattu_2016}.

The prevalence of bias in ML systems has drawn significant attention from researchers in both software engineering and machine learning communities, as it has the potential to cause harm to society. 
To address this issue, various efforts have been made to uncover and mitigate bias in ML systems.
For example, Sun et al.~\cite{10.1145/3377811.3380420} have made progress in uncovering and improving bias in commercial machine translation systems. 
Ribeiro et al.~\cite{checklist} have proposed an evaluation framework for language models, which goes beyond simply assessing accuracy and considers the fairness of the models. Recently, Chen et al.~\cite{chen2022maat} introduce MAAT, an ensemble approach that strikes a balance between fairness and performance in ML software. 
While MAAT achieves state-of-the-art performance, it has limitations to be addressed. 
In particular, it does not take into account the biasness of each example and conducts under-sampling by randomly removing examples.
However, each training example has a varying degree of bias to the final results of a model. We argue that fairness-oriented model building should be achieved through sampling in accordance with data bias rather than randomly.

To mitigate the limitations of existing approaches, we introduce \textit{Counterfactual Fair Software Approach (CFSA)}, a novel framework that aims to improve the fairness and performance of ML software by identifying and removing the bias encoded in the training data. Unlike existing methods that treat all training examples in a subgroup equally, our framework adopts the concept of counterfactual fairness of a training example to quantify its impact on the fairness of the model.
Based on this concept, we construct a ranked \textit{counterfactual bias list}, where examples with a higher rank are more likely to introduce bias into the model. Intuitively, these examples should be removed from the training data with higher priority. Our framework also includes a fair data synthesis approach to maintain class balance when generating new training examples, avoiding exacerbating data imbalance and ensuring that the synthesized dataset closely resembles the original dataset.

We evaluate our approach and the baseline methods on 8 benchmark datasets that consider multiple types of bias, e.g., race, gender, and age bias.
We compare with baselines from both machine learning and software engineering communities.
Reweighing (REW)~\cite{kamiran2012data}, Adversarial Debiasing (ADV)~\cite{10.1145/3278721.3278779} and Reject Option Classification (ROC)~\cite{6413831} are from machine learning venues and have been integrated into the FAIR260 toolkit maintained by IBM~\cite{bellamy2019ai}.
Fair-SMOTE~\cite{10.1145/3468264.3468537} and MAAT~\cite{chen2022maat} are from software engineering venues.
The investigated methods also cover bias mitigation at different stages, e.g., pre-processing techniques that remove biased examples before training the models.

Our experiment results show that CFSA can effectively identify biased data from a large set of training data.
CFSA is also model-agnostic: its effectiveness can be observed on multiple types of models and datasets we investigate.
We run CFSA for 1,500 times on different configurations. 
We find that it outperforms the state-of-the-art in terms of reducing bias in 80\% cases.
Statistical tests also show that CFSA can provide statistically significant better fairness improvement than the baselines.
We also explore the impact of ensemble strategy on the performance of CFSA.
After removing biased examples, we train a model on the new dataset and obtain a \textit{fair} model.
We also train a model on the original dataset, which is called the \textit{performance} model.
We find that the previous settings using in MAAT~\cite{chen2022maat} to ensemble outputs from the two models is not the optimal setting.
Giving larger weights to the outputs from the performance model can further balance the trade-off between accuracy and fairness of the model.
Our additional analysis also demonstrates that CFSA is capable of reducing bias for multiple sensitive attributes simultaneously.

We make the following contributions in the paper:

\begin{itemize}
    \item We propose CFSA, a novel bias mitigation approach that can improve the fairness and performance of ML software by identifying and removing bias encoded in the training data.
    \item CFSA investigates the impact of each training example on the fairness of the model and constructs a ranked list where biased examples have higher ranks. We empirically show that removing examples from the ranked list can better mitigate the bias in the model than the state-of-the-art methods.
    \item We design a fair data synthesis approach that can maintain the class balance when generating new training examples. Experiments show that including the new examples generated by our approach can further improve the fairness and performance of the model.
\end{itemize}

The paper is organized as follows.
Section~\ref{sec:notations} explains some preliminaries and background of this paper.
We explain the details of our methodology in Section~\ref{sec:methodology} and present the experiment settings in Section~\ref{sec:experiment}.
Section~\ref{sec:Result Analysi} describes our answers to various research questions. 
Additionally, related work is discussed in Section~\ref{sec:rel-work}, the conclusion of the paper can be found in Section~\ref{sec:conclusion}.

\section{Preliminaries and Background}
\label{sec:notations}

This section begins by presenting the necessary background information and notation. Next, biases in real-world software development are discussed, showing the root causes of model bias.

\subsection{Terminology}
\label{sec:terminology}

Before we explore the causes of model bias, we provide initial notations and concepts that lead to bias. First, we will review the definition of class label, feature and sensitive attribute.

\begin{itemize}
    \item Class label: A class label indicates the class or category to which an instance belongs to. For example, in a classification problem where the goal is to predict whether an applicant is qualified for the job, ``hire'' and ``do not hire'' are the class labels. In addition, ``hire'' is called favored class label as it gives an advantage to the applicant, while ``do not hire'' is a deprived class label.  
    \item Feature: Feature is an individual's measurable property or characteristic of a phenomenon being observed. It is a specific aspect of the data used as input to a machine-learning model to make predictions.
    \item Sensitive attribute: Sensitive attribute is a special feature that divides the instances into favored and deprived groups. Examples of sensitive attributes include race, gender, age, etc. 
\end{itemize}

We will follow up by introducing the concept of machine learning fairness, which is generally divided into group fairness and individual fairness:

\begin{itemize}
    \item Group fairness: first identifies sensitive attribute(s) that define(s) a potential source of bias then asks for some fairness statistic of the classifier, e.g., prediction accuracy, across favored and deprived groups. 
    \item Individual fairness: requires similarly situated individuals to receive similar probability distributions over class labels to prevent inequitable treatment.
\end{itemize}

In this study, we further consider a particular type of bias embodied in real-world: the individuals' sensitive attribute highly or completely decide their class label showing profound bias towards the deprived groups. To this end, we assume each individual has a corresponding counterfactual individual that only differs in the sensitive attribute in feature space and requires their similar class label prediction.  

With the key concepts outlined, we now introduce the corresponding notation. Let $D = \{d_1, d_2, \dots, d_n\}$ be a dataset, where each $d_i= \{A, S, Y\}$ is described by a set of attributes $A$ with $S$ denoting the sensitive attribute such as gender or race, and the class label $Y$ with $\hat{Y}$ as the predicted class label.  
We also define $D' = \{d_1', d_2', \dots, d_n'\}$ as the counterfactual dataset, where $d_i'$ only differs in $S$ in comparison to $d_i$. Mathematically, $d_i = \{ Y | S = s, A\}$, and $d_i' = \{ Y | S = \overline{s}, A\}$ in which $s\in S$ is a specific value, referred to as the \textit{sensitive value}, e.g., female/black, that defines deprived group (with $\overline{s}$= male/non-black defining favored group). $F$ is a classifier that outputs a predicted probability $F(d_i)= P(\hat{Y}| S, A)$ for each input sample $d_i$.

\subsection{Root of Model Bias}
\label{sec:Root of data bias}

This section explores the root causes of model bias. Previous studies~\cite{chakraborty2021bias,kamiran2018exploiting,kamiran2012data} have demonstrated that classification models constructed using the datasets we used are prone to showing bias. 
One particular cause is that the collected data is often a reflection of the systems and processes that were in place at the time the data was generated. These systems and processes can be influenced by a wide range of factors, including social, cultural, and economic biases. For example, the classification models built from these datasets may be biased if the human labelers had conscious or unconscious biases against certain social groups. This could lead to unfair labeling of the data, which could in turn lead to biased classification models.



Another root cause is the selection bias. For example, a road evaluation program may mistakenly conclude that roads in high-income neighborhoods are worse than in low-income neighborhoods due to selection bias in the data. This was seen in a real-world case where a road maintenance program ``Street Bump", a project by the city of Boston to crowdsource data on potholes~\cite{barocas2017fairness}. The app received more requests from high-income communities due to higher smartphone penetration and user proficiency in those areas. This leads to a model that incorporates political and economic biases.

In the next sections, we delve deeper into the concepts discussed by examining data imbalance and biased data labeling in more detail.

\subsubsection{Data Imbalance Bias}
\label{sec:dataimbalance}

\begin{figure}[!http]
	\centering
	\includegraphics[width=0.45\textwidth]{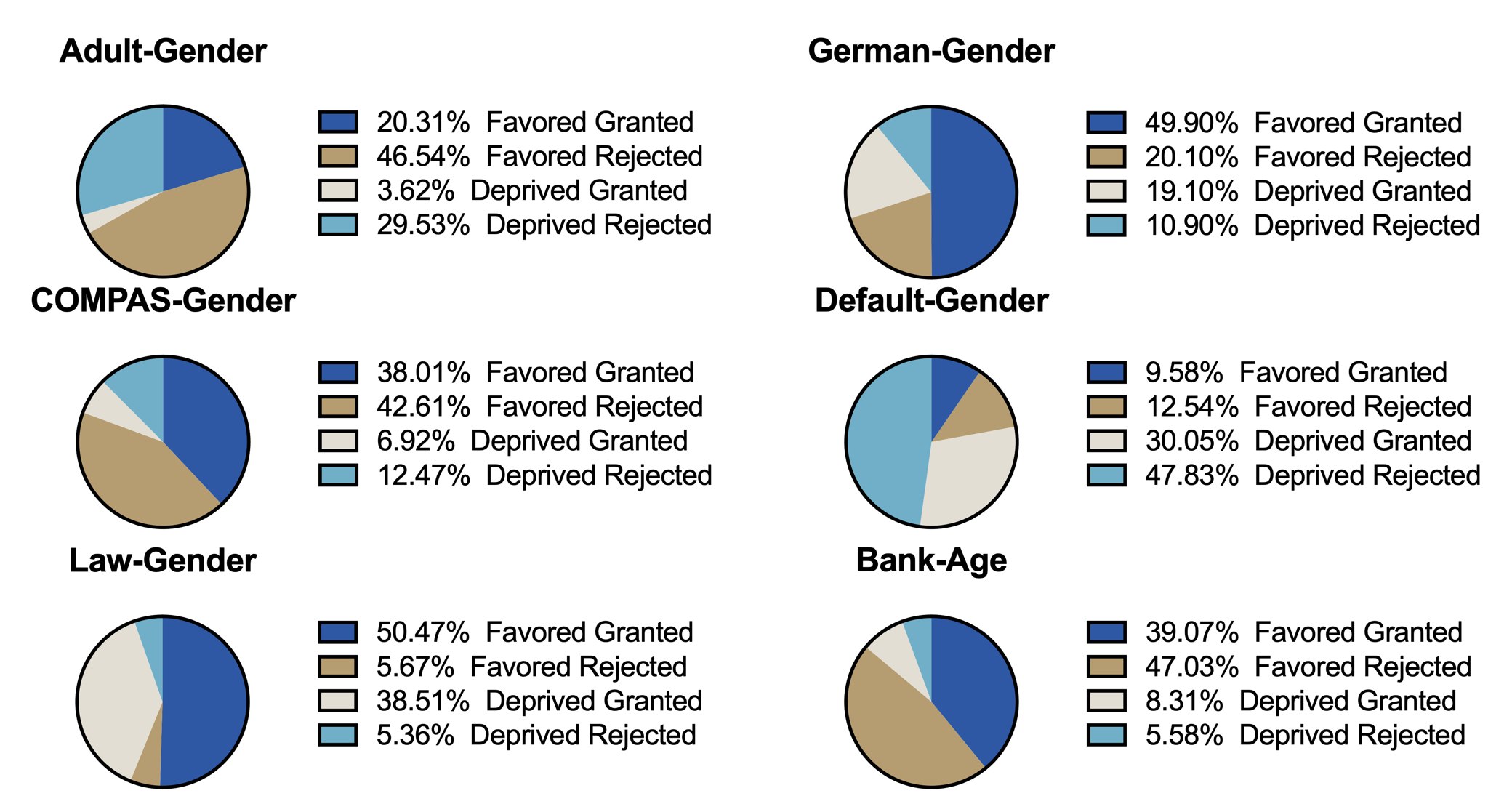}
	\caption{All datasets exhibit an imbalanced distribution concerning the sensitive attribute and class label.}
	\label{fig:toyexample}
\end{figure}

Real-world data sets often have unbalanced distributions, 
as shown in an example in Figure~\ref{fig:toyexample}, the Adult dataset used to predict annual income. In this dataset, high-income individuals make up only 23.93\% of the total, with low-income individuals making up 76.07\%. Additionally, men make up the majority of the high-income category (20.3\%) compared to women (3.62\%), and men also make up the majority of the low-income category (46.54\%) compared to women (29.53\%). This imbalance can affect machine learning models, leading to bias towards assigning favorable labels to male and unfavorable labels to female. The remainder of the datasets also exhibits the same issue.

\subsubsection{Labeling bias}
\label{sec:improperdatalabeling}

In addition to distribution bias, unfair labeling can also lead to bias in machine learning models. Previous research has shown that incorrect labeling can result in individuals with similar characteristics being treated differently. For example, if two samples, one from favored and another from deprived group, have similar features except for the sensitive attribute with distinct class labels, it can be due to the fact that the sample from the deprived group is labeled in a biased way. This can further increase bias towards the deprived group, particularly if those deprived samples are selected for balancing as part of the sampling strategy. To address this issue, we use a counterfactual thinking (c.f., Section~\ref{Counterfactual fairness test and CD List}) to identify samples with biased labels for bias mitigation. 

\section{Methodology}
\label{sec:methodology}



This section proposes solutions to address the aforementioned root causes of model bias toward fair machine learning software development.


\subsection{Briefly}
\label{Briefly}

CFSA is a framework that aims to improve the fairness and performance of ML software by addressing bias at its origin, \textit{i.e.}, the biased encoded in data, while jointly optimizing for performance via ensemble learning. CFSA identifies biased data encoding by evaluating the counterfactual fairness of each sample and creates a \textit{Counterfactual Bias List} (c.f., Section~\ref{Counterfactual fairness test and CD List}) to balance biased representation (c.f., Section~\ref{sec:Data Sampling}) and to correct labeling bias (c.f., Section~\ref{sec:Label introduction bias} and~\ref{sec:synthesisalgorithm}) for fairness improvement. Additionally, it uses ensemble theory to combine models with different optimization objectives to achieve the best results when making predictions (c.f., Section~\ref{sec:performancemodel} and~\ref{sec:combination}). CFSA is a tool that balances fairness and performance of ML software. 

\begin{figure}[!htbp]
	\centering
	\includegraphics[width=0.5\textwidth, height=0.35\textwidth]{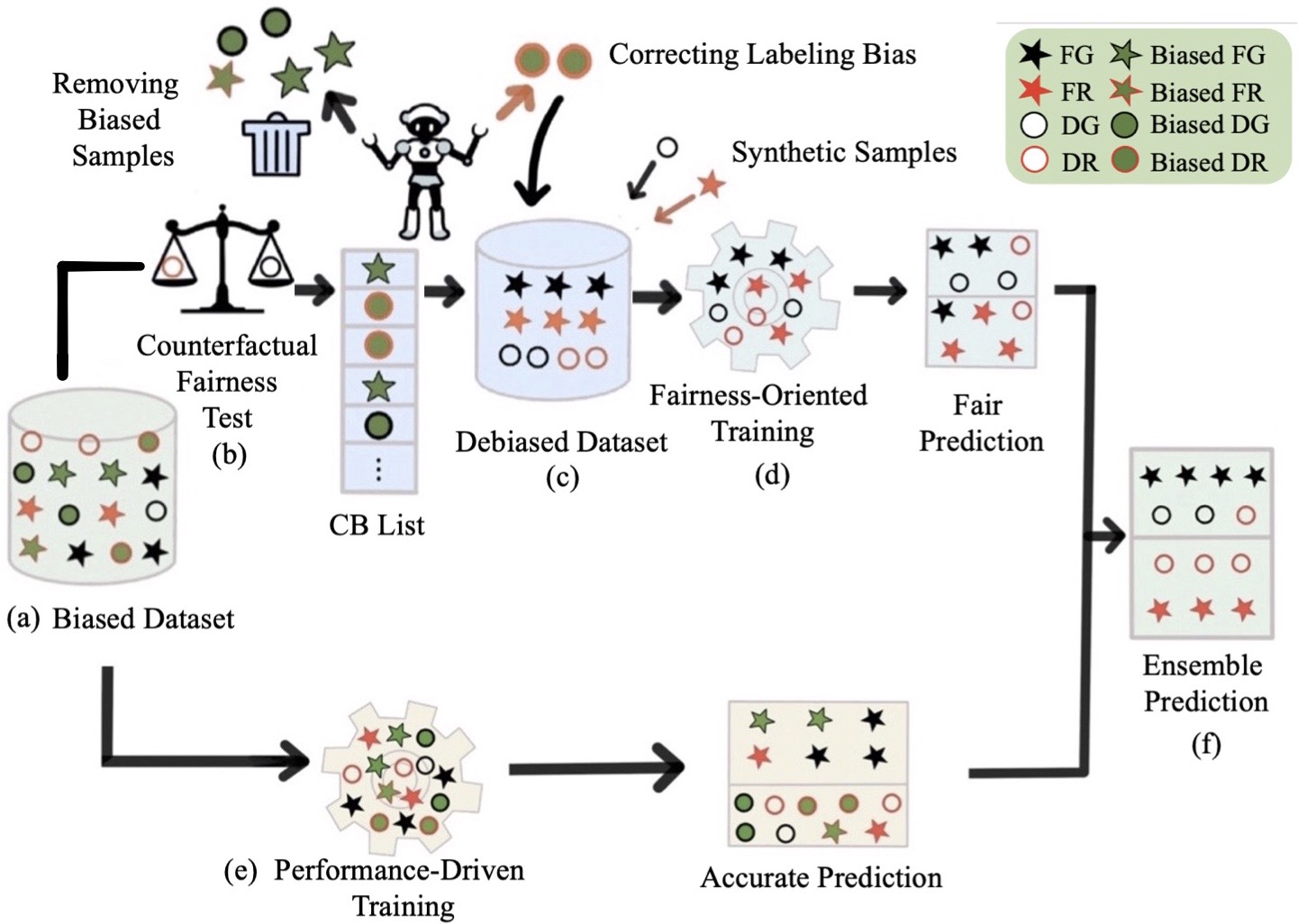}
	\caption{The overall framework of CFSA: (a) Biased dataset; (b) Counterfactual fairness test; (c) Debiased dataset; (d) Fairness-oriented training; (e) Performance-driven training; (f) Ensemble prediction.}
	\label{fig:overview}
	\vspace{-0.5cm}
\end{figure}

\subsection{Counterfactually Debiasing Biased Dataset}
\label{sec:Debiasing Biased Dataset}
\sloppy
As previously discussed in section~\ref{sec:Root of data bias}, bias in the model stems from the dataset. By addressing bias in the dataset, the model and its predictions can also be fairer. Improving the fairness of the model is thus equivalent to improving the fair representation of the dataset, which can be achieved by balancing the representation of different groups and correcting biased labels.


\subsubsection{\textbf{Counterfactual Bias List}}
\label{Counterfactual fairness test and CD List}

The previous works~\cite{DBLP:journals/corr/abs-1901-04966, das2021fairness, chakraborty2021bias} have shown that biased labeling could be a major root behind discrimination. To address this inherent bias, we propose to counterfactually testing whether the decision for a given individual would flip if the value of this individual's sensitive attribute changes. 
As an illustrating example, consider the sensitive attribute $S$ to be gender with the value of $s$ being female ($\overline{s}$ being male) representing deprived community, and the label $Y$ being binary with values of "rejected" or "granted". By combining these two binary features, the dataset $D$ can be divided into four subgroups:

\begin{itemize}
    \item Deprived Rejected ($DR$): Females being rejected a benefit.
    \item Deprived Granted ($DG$): Females being granted a benefit. 
    \item Favored Rejected ($FR$): Males being rejected a benefit.
    \item Favored Granted ($FG$): Males being granted a benefit.
\end{itemize}

Our goal is to test if changing the sensitive attribute $S$ while keeping the insensitive attribute $A$ constant would lead to the flipping prediction of this individual, showing inherent bias toward this deprived community. To this end, we define \textit{Counterfactual Flip Test (CFTest)} of each instance mathematically as:




\begin{equation}
\label{equ:compare}
    CFTest = |F(d_i)>0.5| \oplus |F(d_i')>0.5|
\end{equation}

\noindent where $F(d_i)$ and $F(d_i')$ are the classifier $F$'s predictions, i.e., predicted class probability, on the selected and corresponding counterfactual instances. In addition, $\oplus$ stands for exclusive disjunction, testing whether both sides of the equation are equal. 
By calculating Equation~\ref{equ:compare}, CFTest equals to 1 when the class label flips showing inherent bias and 0 otherwise. Practically, a classifier is learned using all other instances as the training set and then to predict the class labels of the selected instance (i.e., $d_i$) and its corresponding counterfactual instance (i.e., $d_i'$) to calculate CFTest value for each instance. 

The CFTest is strictly defined from class label's perspective and is loosely constrained in terms of predicted probability. Now, let's consider the two following cases from predicted probability's perspective:    






\begin{itemize}
    \item Case 1: $F(d_i)$= 0.9, and $F(d_i')$ = 0.6. 
    \item Case 2: $F(d_i)$= 0.55, and $F(d_i')$ = 0.45. 
\end{itemize}




As we can see, although Case 1 does not show bias according to CFTest, the predicted probabilities differ more significantly in comparison to Case 2, which leads to label flipping but smaller predicted probabilities difference. Therefore, relying on label flipping alone is insensitive to prediction probability based bias. To account for this, we further define \textit{Counterfactual Deviation Test (CDTest)} to measure the prediction probability deviation between the instance and its counterfactual instance as below:



\begin{equation}
\label{equ:FC}
    CDTest = |F(d_i) - F(d_i')|
\end{equation}


We now integrate CFTest and CDTest into a joint objective called \textit{Counterfactual Bias Test (CBTest)}:

\begin{equation}
    CBTest(X_i)= \begin{cases}
        CFTest + CDTest & \text{if CFTest $\neq$ 0} \\\\
        CDTest & \text{Otherwise}
    \end{cases}
\end{equation}



Clearly, CBTest jointly considers the bias rooted in the form of class label flipping as well as prediction distribution. For individuals showing no class label bias towards their counterfactual representation (i.e., FCTest = 0), their CBTest is reduced to CDTest. Based on CBTest values, a ranking list can be created showing the level of bias of each instance. We name such a list as \textit{Counterfactual Bias List (CBList)}, which forms the basis for our following debiasing techniques.

\subsubsection{\textbf{Balancing Biased Representation}}
\label{sec:Data Sampling}

We will now proceed to address imbalance based data bias by utilizing the \textit{CBList}, which serves as the basis to identify individuals that prone to bias for rebalancing data distribution. Before going further into this, we first discuss our guiding principle for addressing bias caused by unbalanced data distributions the \textit{``We are all equal'' (WAE)} worldview~\cite{friedler2021possibility}. Specifically, WAE calls for equal probability of being granted for both deprived and favored communities, mathematically represented as:


\begin{equation}
\label{equ:wae}
    \frac{FG}{FG+FR} = \frac{DG}{DG+DR}
\end{equation}

In real-world data, however, the distribution is often highly imbalanced thus WAE does not hold. As discussed in section~\ref{sec:dataimbalance}, the probability of being granted for favored community will be significantly greater than the probability of deprived community's, i.e., the value of the left side of Equation~\ref{equ:wae} is larger than the right side's, as a result of the relative significant overrepresentation in $FG$ and $DR$. To this end, undersampling is performed on $FG$ and $DR$ to align with the view of WAE:


\begin{equation}
\label{equ:reduce}
    \frac{FG - FG_{Remove}}{FG - FG_{Remove} + FR} = \frac{DG}{DG + DR - DR_{Remove}}
\end{equation}

\noindent where $FG_{Remove}$ and $DR_{Remove}$ are the number of samples to be removed from $FG$ and $DR$, respectively. 

There are various combinations of $FG_{Remove}$ and $DR_{Remove}$ to achieve this goal. To mitigate bias while preserving as much original data information as possible, the combination that maintains the original relative representation between favored and deprived group is desired:


\begin{equation}
\label{equ:ratiosame}
    \frac{FG + FR}{DG + DR} = \frac{FG_{Remove}}{DR_{Remove}}
\end{equation}


With the desired $FG_{Remove}$ and $DR_{Remove}$ determined based on Equation~\ref{equ:reduce} and Equation~\ref{equ:ratiosame}, we utilize \textit{CBList} to determine exactly which individuals to be removed. Specifically, the top $FG_{Remove}$ and $DR_{Remove}$ ranked individuals in \textit{CBList} from $FG$ and $DR$ are undersampled respectively to meet the criteria of WAE (further bias correction operation on undersampled $DR$ is discussed in the following Section~\ref{sec:Label introduction bias}). Using this debiasing guided sampling, as opposed to the random sampling methods used in previous studies~\cite{chen2022maat, 10.1145/3468264.3468537}, can effectively lead to a procedure for balancing that mitigates bias given the bias identification power of \textit{CBList}.

\subsubsection{\textbf{Correcting Labeling Bias}}
\label{sec:Label introduction bias}

With a balanced representation, we further address labeling bias also on the basis of the proposed \textit{CBList}. Since labeling bias in $FG$ and $DR$ has been addressed in the previous debiasing balancing procedure by selecting individuals with labeling bias for undersampling, we focus on $FR$ and $DG$ at this stage. In specific, all individuals from $FR$ and $DG$ that do not meet the criteria of counterfactual fairness are removed; removing individuals from $FR$ and $DG$ with CBTest values greater than 1 which indicates that the classifier produces different predictions in the real and counterfactual worlds. 

In addition, to ensure WAE still holds, corresponding number of removed instances will be added back through either synthesization or class label flipping. First, regarding $DG$, relating to the aforementioned bias correction for undersampled individuals in $DR$ in Section~\ref{sec:Data Sampling}. Specifically, their discriminatory treatment due to the presence of their sensitive attribute is a major manifestation of bias in real-world. Corresponding to \textit{CBList}, such a manifestation is equivalent to the flipping class label solely due to one individual's sensitive attribute, \textit{i.e., CFTest}. To correct these biased labels, we flip such $DR$ individuals' class label and re-include them as $DG$ in the dataset to mitigate bias while preserving as much original data as possible. In the event that the number of removed individuals is greater than the number of flipped individuals, corresponding number of instances will be synthesized (c.f., Section~\ref{sec:synthesisalgorithm} for synthesizing details) and included in $DG$. Second, in terms of $FR$, synthesization is applied directly without class label flipping as $FR$ is typically not the major manifestation of labeling bias. The data is now prepared for building fairness-oriented models, having undergone balancing of representation and biased label correction.

\subsubsection{\textbf{Fair Synthesis}}
\label{sec:synthesisalgorithm}

Our synthetic algorithm addresses the issue of increased intra-class imbalance present in traditional oversampling techniques such as ROS-random oversampling, SMOTE and KMeans-SMOTE~\cite{last1711oversampling}. It is designed to maintain the class balance while generating new samples, thus avoiding exacerbation of the in-class imbalance. This is achieved through a three-step process that includes clustering, filtering, and synthesizing: i) the algorithm first groups the data based on sensitive attribute and class label, then each subgroup is clustered using k-nearest neighbors~\cite{peterson2009k}, thus individuals with similar characteristics will be grouped together within subgroups. ii) To avoid blurring of the clustering boundaries, the algorithm filters each class by removing the farthest 20 percent of points from the center of the clusters from the sample points. iii) Finally, the algorithm generates simulated data proportionally in the different classes to ensure that the distribution of the synthesized dataset is consistent with that of the original dataset. This approach avoids exacerbating the intra-class imbalance while generating new samples, making the synthesized dataset more representative of the original data.

\subsection{Accuracy-Driven Training}
\label{sec:performancemodel}


In CFSA, the objective of the performance model is to maximize its performance. To achieve this, various machine learning (ML) algorithms are used to train the models on the original training data $D$. The model with the highest accuracy is chosen as the performance model. In the experimental analysis, the effect of using different performance models on the final results will be explored. This approach allows us to determine the optimal performance model for the given dataset and task.

\subsection{Ensemble Training}
\label{sec:combination}

With the fairness and performance model trained on debiased and original dataset respectively, CFSA now combines the outputs of them to make the final prediction. This involves ensembling the prediction probability vectors generated by each model using the formula outlined in Equation~\ref{equ:Combination}:

\begin{equation}
    \label{equ:Combination}
        \hat{Y} = \sum_{i=1}^{n} W_i \times P_i
\end{equation}


\noindent where $W_i$ is the weight of different models and $P_i$ is the prediction probability vector. Consider the binary classification task with average ensemble as an example, the combination module first takes the prediction probability vector, 
from these two models as inputs, e.g., $P_{f}$ and $P_{a}$ for fairness-oriented training model and accuracy-driven training model, respectively. Next, the combination module uses an averaged weighting strategy, i.e., the weighting vector $W$=[0.5, 0.5], and the final combined prediction probability vector is thus computed as: [$0.5 \times (P_{f}(\hat{Y}=0)+P_{a}(\hat{Y}=0)), 0.5\times(P_{f}(\hat{Y}=1) + P_{a}\times(\hat{Y}=1)$]. When $0.5\times(P_{f}(\hat{Y}=0) + P_{a}(\hat{Y}=0))$ > $0.5\times(P_{f}(\hat{Y}=1) + P_{a}(\hat{Y}=1))$, the label is predicted as 0 and 1 otherwise. In addition to the commonly used averaging strategy in ensemble learning, we will also explore other combination strategies in Section~\ref{sec:Result Analysi} to investigate how different strategies affect the validity of our results which will shed light on finding the best combination strategy for the given dataset and task.


\section{Experiment Settings}
\label{sec:experiment}

In this section, we describe our experimental setting and experimental datasets to answer the research questions in Section~\ref{sec:Result Analysi}. We first describe the datasets used in our experiments and then present the baselines and metrics selected in our experiments. 


\subsection{Datasets}
\label{sec:datasets}



In contrast to typical fair machine learning studies that only evaluate up to three datasets~\cite{hort2021fairea, zhang2019faht, zhang2021farf, zhang2021fair}, our method is comprehensively evaluated on eight real-world datasets with varying feature spaces and sensitive attributes (the details of the datasets used can be found in in Table~\ref{tab:dataset_info}), covering various domains as follows:



\textit{Financial domain}:
i) The \emph{Adult} dataset~\cite{fox2012rcmdrplugin} is used for a prediction task aimed at determining whether a person earns an annual income of over \$50K based on their demographic and financial information. 
ii) The \emph{Bank} dataset~\cite{moro2014data} is employed for predicting whether a client of a bank will opt for a term deposit, based on their demographic, financial, and social information. 
iii) The \emph{German} dataset~\cite{dheeru2017uci} is a financial dataset of bank account holders, commonly used for predicting creditworthiness to assess credit risk.  
iv) The \emph{Default} dataset~\cite{yeh2009comparisons} studies the default payments of customers in Taiwan, with the aim of predicting the probability of default in the next month.

\textit{Criminological domain}: v) The \emph{COMPAS} dataset~\cite{angwin2016machine} is a well-known dataset in the field of algorithmic bias, used for predicting the likelihood of criminal recidivism in defendants.

\textit{Social domain}: vi) The \emph{Dutch} dataset~\cite{van20002001} compiles information on individuals in the Netherlands for the year 2001 and is utilized for predicting a person's occupation. 
vii) The \emph{Heart} dataset~\cite{tarawneh2019hybrid}, dating back to 1988, gathers medical information on patients and is employed for predicting the presence or absence of heart disease. 

\textit{Educational domain}: viii) The \emph{Student} dataset~\cite{wightman1998lsac} contains law school admission records and is used to predict if a candidate will pass the bar exam and their first-year average grade. 




\begin{table}[!htbp]
 	\caption{Summary of datasets used in experiments.}
	\small
	\begin{tabular}{|c|c|c|c|c|}
		\hline
		\diagbox{Dataset}{CHAR}  & Sample\# & Features\# & \makecell[c]{Sensitive \\ Attribute} \\
		\hline
		Adult  			& 45,222       & 12      & Race,Gender   \\
		Bank 		    & 45,211       & 17      & Age   \\
        COMPAS  		& 6,172        & 12      & Race,Gender \\
		German  		& 1,000        & 21      & Gender \\
        Default  		& 30,000       & 11      &  Gender   \\
		Dutch 		    & 60,420       & 12      & Gender   \\
		Law  			& 20,798       & 12      & Gender \\
        Heart 		    & 297       & 14      & Age   \\
		\hline
	\end{tabular}  
	\label{tab:dataset_info}
\end{table}

\subsection{Baselines}
\label{sec:Baseline}
To evaluate the performance of our method, we compare five existing bias mitigation methods: Reweighing (REW)~\cite{kamiran2012data}, Adversarial Debiasing (ADV)~\cite{10.1145/3278721.3278779}, Reject Option Classification (ROC)~\cite{6413831}, Fair-SMOTE~\cite{10.1145/3468264.3468537}, and MAAT~\cite{chen2022maat}. The first three baselines are recent advanced methods proposed by the ML community, which are integrated into the IBM AIF360 toolkit~\cite{bellamy2019ai}. The remaining two baselines are state-of-the-art approaches recently proposed in the Software Engineering venues. These baselines cover a wide range of debiasing methods including pre-processing, in-processing, post-processing, and ensemble. Next, we briefly describe each approach.

\begin{itemize}
    \item The REW~\cite{kamiran2012data}, Fair-SMOTE~\cite{10.1145/3468264.3468537}, and MAAT~\cite{chen2022maat} are pre-processing methods used to address bias in machine learning algorithms: i) REW calculates the weight of each group based on the label and protection attributes. ii) Fair-SMOTE uses a combination of data clustering and oversampling to equalize the number of training data in different subgroups. iii) MAAT divides the dataset into four groups and adjusts the sample size of each group to ensure equal favorable rates for favored and deprived groups, then fair and accurate models are trained and combined through an ensemble method.
    \item ADV~\cite{10.1145/3278721.3278779} is an in-processing method for addressing bias in machine learning algorithms. It uses adversarial techniques to reduce the impact of sensitive attributes on the model's predictions while maximizing overall performance.
    \item ROC~\cite{6413831} is a post-processing method that reduces bias in machine learning algorithms. It focuses on predictions with high uncertainty and reassigns them to reduce bias. Specifically, it aims to allocate favorable outcomes to deprived groups and unfavorable outcomes to favored groups.
\end{itemize}

\subsection{Evaluation Metrics}
\label{sec:Metrics}


The evaluation involves three fairness metrics and five ML performance metrics. We will first present the fairness measures, then the ML performance metrics, and finally describe how to quantify the trade-off between fairness and performance.

\subsubsection{Fairness Metrics}
\label{sec:FairnessMetrics}
To measure ML software fairness, we employed three commonly used fairness metrics: Statistical Parity Difference (SPD), Average Odds Difference (AOD), and Equal Opportunity Difference (EOD). Our choice of these metrics is based on their widespread adoption in the field, as demonstrated in the literature~\cite{spinellis2021proceedings}.

\begin{itemize}
    \item SPD: quantifies the disparity between the probability of the favored and deprived group receiving a benefit:
    
    \begin{equation}
        SPD = P\left[\hat{Y} = 1| S = 0\right] - P\left[\hat{Y} = 1| S = 1\right]
    \end{equation}

    \item AOD: measures the average of the False Positive Rates (FPR) and the True Positive Rates (TPR) between favored and deprived group:
    
    \begin{equation}
    \begin{aligned}
        AOD = \frac{1}{2}(\left|P\left[\hat{Y} = 1| S = 0, Y = 0\right]- P\left[\hat{Y} = 1| S = 1, Y = 0\right]\right| \\
        + \left|P\left[\hat{Y} = 1| S = 0, Y = 1\right] - P\left[\hat{Y} = 1| S = 1, Y = 1\right]\right|)
    \end{aligned}
    \end{equation}
    
    \item EOD: measures the True Positive Rates (TPR) difference between favored and deprived group:
    
    \begin{equation}
        EOD = P\left[\hat{Y} = 1| S = 0, Y = 1\right] - P\left[\hat{Y} = 1| S = 1, Y = 1\right].
    \end{equation}
\end{itemize}


We use the absolute value of all fairness metrics, with zero representing maximum fairness and higher values indicating greater bias.

\subsubsection{Performance Metrics}
\label{sec:PerformanceMetrics}


To measure ML software performance, we employed five performance metrics: accuracy, recall, Matthews correlation coefficient, precision, and F1-Score. These metrics can be calculated using the confusion matrix of a binary classification, which comprises of four elements: true positives (TP), false positives (FP), true negatives (TN), and false negatives (FN).

\begin{itemize}
    \item Accuracy: measures the fraction of correct predictions made by the model out of all the predictions.
    \begin{equation}
        Accuracy = \frac{TP+TN}{TP+FP+TN+FN}
    \end{equation}
    
    \item Recall: measures the proportion of actual positive cases that the model correctly identified.
    \begin{equation}
        Recall = \frac{TP}{TP + FN}
    \end{equation}
    
    \item Matthews Correlation Coefficient(MCC): measures the different between true positive and true negative rate.
    \begin{equation}
        MCC = \frac{TP \times TN - FP \times FN}{\sqrt{(TP + FP)(TP + FN)(TN + FP)(TN + FN)}}
    \end{equation}
    
    \item Precision: refers to the fraction of positive predictions that are actually correct.
    \begin{equation}
        Precision = \frac{TP}{TP+FP}
    \end{equation}
    
    \item F1-Score: represents the harmonic mean of precision and recall.
    \begin{equation}
        F1-Score = \frac{2 \times (Precision \times Recall)}{(Precision + Recall)}
    \end{equation}
\end{itemize}


For all of the performance metrics, the higher the value the better performance. In addition, the value of all metrics other than MCC, i.e., accuracy, recall, precision, and F1-Score, range from 0 to 1. In terms of MCC, it has a range of -1 to 1, where 1 represents a perfect forecast, 0 represents a prediction that is no better than random, and -1 indicates a prediction that completely contradicts the observation. Among all the performance metrics, MCC also is more sensitive to the overall quality of the model's predictions as it takes both the true and false positive and negative rates into account, and has been demonstrated to be suitable for dealing with imbalance in various software engineering research~\cite{chicco2020advantages}.

\subsubsection{Jointly evaluating Fairness and Performance of the Model}
\label{sec:measurements}



Based on the aforementioned fairness and performance metrics, the benchmarking tool \textit{Fairea}~\cite{hort2021fairea} is employed to jointly evaluate the effectiveness of fairness and performance of the ML software models. The fundamental idea of Fairea is to convert the original model into a random guessing model to enhance its fairness. This is achieved by mutating predictive class label such that predictive performance are equally worse in both favored and deprived group. Therefore, it's anticipated that effective bias mitigation techniques will surpass the fairness-performance trade-offs of mutated models. Specifically, it operates in the following manner: 

\textit{Trade-off baseline}: Fairea begins by utilizing the initial model to make predictions and then duplicates these predictions. Next, \textit{Fairea} randomly selects predictions and mutates these predicted class labels, i.e., changing all of them with the majority class of data, based on different mutation degrees, e.g., 10\%, 20\%, \dots, 100\%. This mutated model is called the pseudo model. In addition, as the mutation degree increases, the accuracy of the model's predictions decreases, but the fairness of the predictions improves as the prediction becomes more random and similar across subgroups. Particularly, when the mutation degree reaches 100\%, all predictions receive the same prediction, resulting in the lowest accuracy but the highest fairness. \begin{wrapfigure}{r}{0.2\textwidth}
	\centering
	\includegraphics[width=0.2\textwidth]{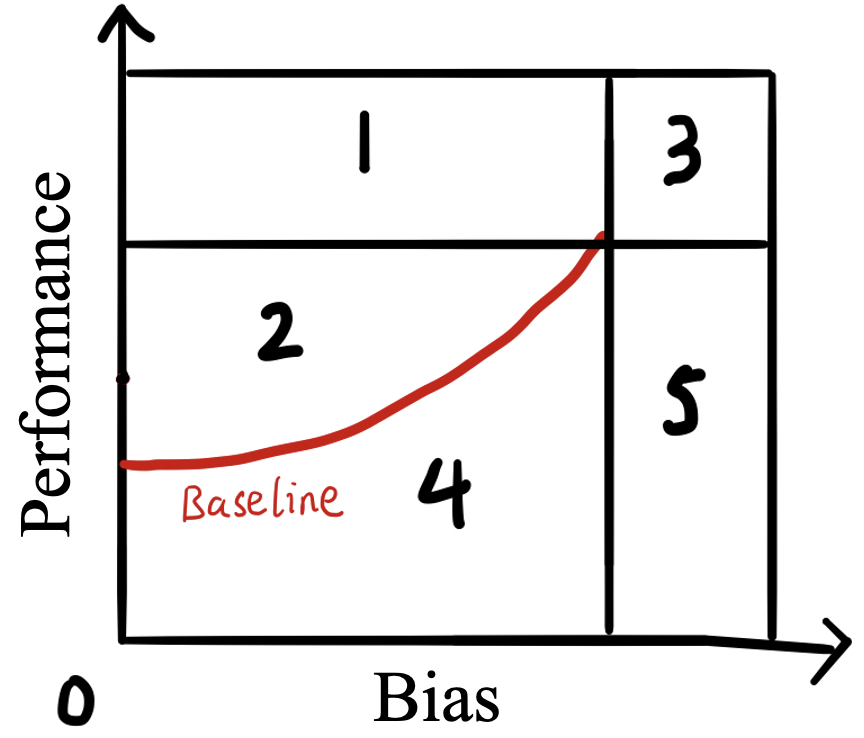}\vspace{-0.3cm}
	\caption{Fairea's mitigation regions based on changes in performance and bias. \vspace{-0.3cm}}
	\label{fig:scalability}
\end{wrapfigure}Lastly, \textit{Fairea} constructs a trade-off baseline for one specific model, e.g., CFSA or other competition baselines, by connecting the fairness-performance (measured by the aforementioned metrics) points of the original model and a series of pseudo models (as show in Figure~\ref{fig:scalability}).

\textit{Five effectiveness levels}: The trade-off baseline categorizes bias mitigation methods into five trade-off effectiveness levels: i) ``win-win'': A method falls into this trade-off if it enhances both ML performance and fairness compared to the trade-off baseline. Such a fairness-performance point will be located in region 1. ii) A method is considered a "good" trade-off if it improves either machine learning performance or fairness compared to the trade-off baseline, and is overall better than the trade-off baseline, thus situating in region 2. iii) If a method improves ML performance but leads to fairness drop, it falls into a ``inverted'' trade-off, being in the region 3. vi) If a method falls into a ``pool'' trade-off its either ML performance or fairness declines compared to the trade-off baseline, and overall worse than the trade-off baseline. The fairness-performance point thus will be located in region 4. v) A method is considered a ``lose-lose'' trade-off if it results in a decline in both machine learning performance and fairness compared to the original model. This results in the fairness-performance point being positioned in region 5.


    

\section{Experimental Results}
\label{sec:Result Analysi}

\subsection{Experimental Setup}
\label{sec:ExperimentalSetup}

The experiment involves the use of 8 datasets, which are shuffled and divided into 70\% training and 30\% test data. Samples with missing values are removed, continuous features transformed into categorical categories, non-numerical features converted to numerical values, and all feature values normalized to [0,1]. Three standard machine learning models, e.g., \textit{Logistic Regression (LR)}, \textit{Support Vector Machine (SVM)}, and \textit{Random Forest (RF)}, are employed as the base model to build our proposed CFSA as well as baselines, e.g., REW, ADV, ROC, and Fair-SMOTE and MAAT. Among them, the first three implementations are based on IBM AIF360~\cite{bellamy2019ai} while the remaining are released by their authors~\cite{10.1145/3468264.3468537,chen2022maat}. 

To implement benchmarking tool Fairea, we establish a trade-off baseline based on each benchmark task, ML algorithm, and fairness-performance measurement. Specifically, we train the original model 50 times. For each trained original model, the mutation is repeated 50 times and each time with a different mutation degree. The trade-off baseline is then constructed using the averaged result from the multiple runs.

The experiments are implemented with Python 3.7 and executed on a 64-bit machine with a 10-core processor (i9, 3.3GHz), 64GB memory with GTX 1080Ti GPU. The experimental results are organized to answer the following five research questions (RQs).


\subsection{Research Questions}
\label{sec:researchquestion}

\paragraph{\textbf{RQ1: Can CBList effectively identifies Biased data samples?}}
\label{RQ:RQ1}



\begin{figure}[!htbp]
	\centering
	\includegraphics[width=0.45\textwidth, height=0.30\textwidth]{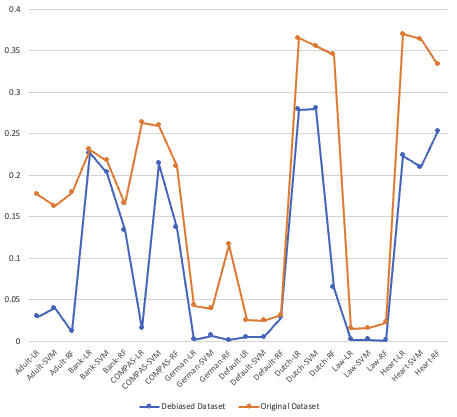}
	\caption{The statistical parity differences with and without biased sampls identified by CBList removed.}
	\label{fig:RQ1}
\end{figure}

This research question focus on understanding the effectiveness of the method in identifying biased samples, which is crucial as addressing bias in the dataset prior to training the model can significantly reduce the likelihood of biased decisions. To answer this question, we continue to use the three aforementioned machine learning models as the base model to construct the CBList. Based on this, instances prone to bias are identified then removed for fairness-oriented training. The results, as shown in Figure~\ref{fig:RQ1}, indicate that all models trained on the original biased datasets (represented by the yellow line) have higher bias scores (only statistical parity difference is shown for the ease of distinction) than the models trained on the CBList debiased datasets (represented by the blue line), with the largest fairness improvement by 98.05\% in the Law dataset. This suggests that CBList can effectively identify biased data instances for fairness-oriented model training. Additionally, there is no notable result difference across three base models built in conjunction with CBList, suggesting that CBList is model agnostic for fair ML software tasks.

\textbf{Ans. to RQ1:} Yes, the CFSA improves fairness by as much as 98.5\%. To conclude, CBList is effective in identifying biased instances, and having these instances removed exactly mitigated the biased data representation.

\begin{figure}[!htbp]
	\centering
	\includegraphics[width=0.45\textwidth, height=0.2\textwidth]{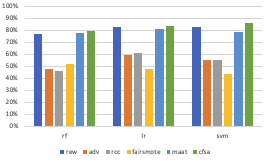}
	\caption{Proportion of cases where CFSA beats the baseline in different ML algorithms.}
	\label{fig:RQ2}
\end{figure}

\paragraph{\textbf{RQ2: Can CFSA reduce bias?}}
\label{RQ:RQ2}
To answer this question, we first evaluate the effectiveness of CFSA in 10 uni-attribute benchmark tasks. For each task, CFSA is applied with the same three ML models, e.g., LR, SVM, RF, for 50 times in Fairea. Hence, we have $10 \times 3 \times 50 = 1,500$ cases in total. As can be seen in Figure~\ref{fig:RQ2}, CFSA (Green bar) beats the trade-off baseline constructed by mutated CFSA in 81\% of the cases. 


In addition, the reduction in model bias is only considered meaningful if it does not result in a significant decrease in model performance, which can be reflected by the cases where CFSA outperforms its trade-off baseline as shown in rows 6, 12, and 18 of Table~\ref{tab:RQ2_1}. As can be seen, CFSA wins at least 78\% of the cases showing significant bias reduction while maintaining competitive performance.

\begin{table}[]
    \centering
    \caption{Proportions of scenarios where each method significantly improves fairness and decreases performance. CFSA significantly improves fairness in 98.4\% of the scenarios, without decreasing ML performance too much.}
    \begin{tabular}{|c|c|c|c|c|c|c|}
    \hline
         & REW & ADV & ROC & \makecell[c]{Fair-\\SMOTE} & MAAT & CFSA \\
         \hline
        \makecell[c]{Fairness \\ (+)} & 57.1\% & 63.5\% & 90.5\% & 66.7\%&77.8\% &98.4\% \\
        \hline
        \makecell[c]{Accuracy \\ (-)} & 10.5\% & 13.4\% & 25.7\% & 17.1\% & 17.2\%& 22.3\% \\
    \hline
    \end{tabular}
    \label{tab:RQ2}
\end{table}

\begin{table}
    \centering
        \caption{The proportion of mitigation cases that surpass the trade-off baseline in 15 fairness-performance evaluations from CFSA and existing methods (the darker cells show top rank and the lighter cells show the second rank).}
    \begin{tabular}{|c|c|c|c|c|c|}
    \hline
    Methods &\makecell[c]{SPD \\ Accuracy }&	\makecell[c]{SPD \\ Precision }&	\makecell[c]{SPD \\ Recall }&	\makecell[c]{SPD \\ F1Score }&	\makecell[c]{SPD \\ MCC }\\
    \hline
    rew & \cellcolor{blue!15}0.71&	\cellcolor{blue!15}0.74&	\cellcolor{blue!15}0.79&	\cellcolor{blue!15}0.80&	\cellcolor{blue!15}0.78\\
    \hline
adv&	0.69&	0.66&	0.70&	0.70&	0.70\\
\hline
roc&	0.58&	0.51&	0.72&	0.71&	0.72\\
\hline
Fair-SMOTE&	0.32&	0.29&	0.34&	0.34&	0.34\\
\hline
MAAT&	\cellcolor{blue!15}0.71&	0.70&	0.71&	0.73&	0.72\\
\hline
CFSA&	\cellcolor{blue!40}0.81&	\cellcolor{blue!40}0.85&	\cellcolor{blue!40}0.85&	\cellcolor{blue!40}0.84&	\cellcolor{blue!40}0.85\\
\hline
    Methods &\makecell[c]{AOD \\ Accuracy }&	\makecell[c]{AOD \\ Precision }&	\makecell[c]{AOD \\ Recall }&	\makecell[c]{AOD \\ F1Score }&	\makecell[c]{AOD \\ MCC }\\
    \hline
    rew& 0.70&	0.75&	\cellcolor{blue!15}0.79&	\cellcolor{blue!15}0.80&	\cellcolor{blue!15}0.79\\
    \hline
adv& 0.40&	0.38&	0.40&	0.40&	0.40\\
\hline
roc&0.38&	0.38&	0.50&	0.49&	0.50\\
\hline
Fair-SMOTE&0.46&	0.39&	0.47&	0.47&	0.47\\
\hline
MAAT& \cellcolor{blue!15}0.72&	\cellcolor{blue!15}0.76&	0.77&	0.78&	0.77\\
\hline
CFSA& \cellcolor{blue!40}0.80&	\cellcolor{blue!40}0.87&	\cellcolor{blue!40}0.84&	\cellcolor{blue!40}0.85&	\cellcolor{blue!40}0.87\\
\hline
    Methods &\makecell[c]{EOD \\ Accuracy }&	\makecell[c]{EOD \\ Precision }&	\makecell[c]{EOD \\ Recall }&	\makecell[c]{EOD \\ F1Score }&	\makecell[c]{EOD \\ MCC }\\
    \hline
    rew&\cellcolor{blue!15}0.77&	0.76&	0.79&	0.79&	0.79\\
    \hline
adv&0.39&	0.37&	0.40&	0.40&	0.40\\
\hline
roc&0.32&	0.31&	0.42&	0.41&	0.42\\
\hline
Fair-SMOTE&0.56&	0.49&	0.57&	0.57&	0.57\\
\hline
MAAT&0.75&	\cellcolor{blue!15}0.84&	\cellcolor{blue!15}0.85&	\cellcolor{blue!15}0.87&	\cellcolor{blue!40}0.86\\
\hline
CFSA&\cellcolor{blue!40}0.78&	\cellcolor{blue!40}0.86&	\cellcolor{blue!40}0.86&	\cellcolor{blue!40}0.88&	\cellcolor{blue!15}0.85\\
    \hline
    \end{tabular}

    \label{tab:RQ2_1}
\end{table}

\textbf{Ans. to RQ2:} CFSA achieves a good or win-win trade-off in 85\% of the cases while poor or lose-lose trade-off is only 2\%. In sum, CFSA can reduce bias while not resulting in a significant decrease in model performance. 


\paragraph{\textbf{RQ3: How well does CFSA perform compared to the state-of-the-art bias mitigation algorithms?}}

To answer this question, CFSA is evaluated against 5 state-of-the-art baselines, e.g., REW, ADV, ROC, Fair-SMOTE, MAAT, in 10 benchmark tasks. Same as previous RQs, for each task, CFSA and other baselines are constructed using the same 3 base models for 50 times, and each individual run is treated as a distinct mitigation case. As a result, we have total $10 \times ( 1 + 5 ) \times 3 \times 50 = 9,000$ cases. To simplify the demonstration, we use the percentage of mitigation scenarios that exceed the trade-off baseline established by Fairea as a measure of effectiveness (i.e., scenarios that result in either a good or win-win trade-off).

\begin{figure}[!htbp]
	\centering
	\includegraphics[width=0.40\textwidth, height=0.35\textwidth]{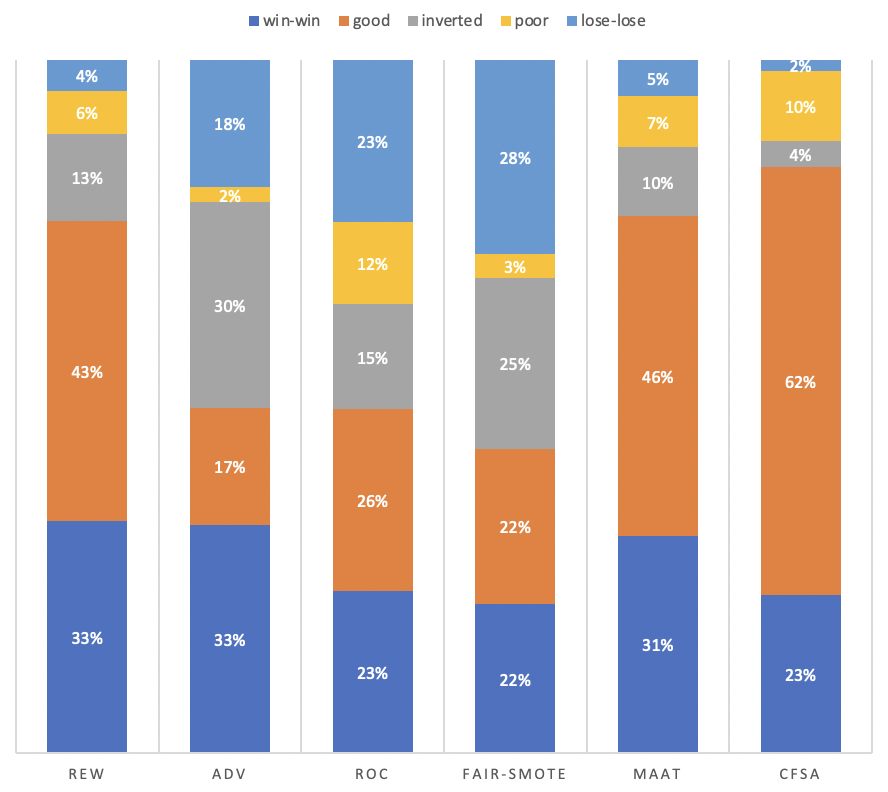}
	\caption{CFSA and other methods' effectiveness distribution in benchmark tasks; CFSA shows the best balance, with 85\% of mitigation cases having good or win-win results.}
	\label{fig:RQ3-1}
\end{figure}

\begin{figure*}[!htbp]
	\centering
	\includegraphics[width=1\textwidth, height=0.20\textwidth]{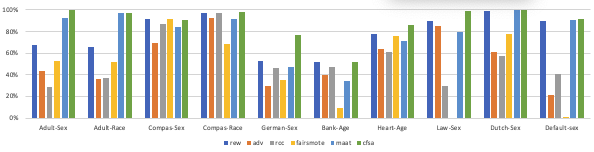}
	\caption{The comparison of trade-off baseline performance between CFSA and 5 baselines.}
	\label{fig:RQ3-2}
\end{figure*}

 Figure~\ref{fig:RQ3-1} shows the overall results. As we can see, CFSA achieves a good or win-win trade-off (i.e., beat the trade-off baseline constructed by Fairea) in most cases, i.e., 85\% of the time. In comparison, the corresponding percentages for REW, ADV, ROC, Fair-SMOTE, and MAAT were 76\%, 50\%, 49\%, 44\%, and 77\%, respectively. In addition, CFSA has significantly fewer lose-lose trade-off cases (2\%) than other existing methods, such as Fair-SMOTE which has a lose-lose trade-off rate of 28\% that is 14 times higher than CFSA. 
 The Figure~\ref{fig:RQ3-2} displays the results more clearly by showing the percentage of times that CFSA and other baselines beat their correpsonding trade-off baseline in 10 benchmark tasks. The results show that the performance of existing methods, such as REW and Fair-SMOTE, is inconsistent across different decision-making tasks. For example, in the Compas-Sex task, REW and Fair-SMOTE outperform the baseline by 97.4\% and 91.7\%, respectively, but in the Adult-sex task, their success rates dropped to 67.5\% and 52.8\%. On the other hand, CFSA show consistent performance, with a success rate of 91.1\% in the Compas-sex task and 99.9\% in the Adult-sex task, with only a small difference of 8.8\%. This highlights the improved performance of CFSA compared to other existing methods in achieving a trade-off between fairness and performance.
 
Additionally, for each combination of task, base models, fairness-performance metric, we compare the percentage of surpassing trade-off baseline of CFSA and five other baselines. The results are displayed in Table~\ref{tab:RQ2_1}, where CFSA, in 15 fairness-performance measurements, secures 14 first place finishes and only 1 second place finish (with a margin of only 1\% to the 1st place).

\textbf{Ans. to Q3:} CFSA achieves the best trade-off, CFSA outperforming other methods at least 8\% more in good or win-win trade-off. Also, CFSA achieves less poor or lose-lose trade-off than other methods. In summary, the superiority of CFSA over state-of-the-art is maintained across all studied ML algorithms, uni-attribute benchmark tasks, and fairness-performance evaluations.



\paragraph{\textbf{RQ4: How do various combination strategies impact the performance of our method?}}
\label{RQ:RQ4}

\begin{figure}[!htbp]
	\centering
	\includegraphics[width=0.35\textwidth, height=0.15\textwidth]{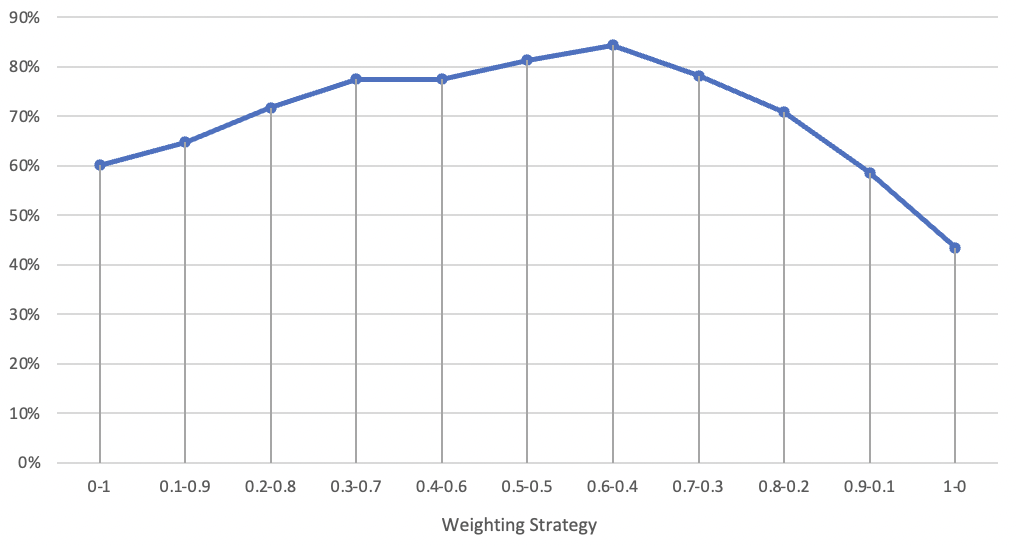}
	\caption{The effectiveness of combination strategies of CFSA, with the [0.6-0.4] strategy being the most effective.}
	\label{fig:RQ4}
\end{figure}

To answer this question, we set the 11 different weighting strategies ranging from 0 to 1 with step size 0.1 (i.e., $W$= [0, 1], [0.1, 0.9], $\cdots$, [0.9, 0.1], [1, 0]). When fairness is the sole focus, the weighting strategy $W$ is set to [1, 0] while [0, 1] purely focus on performance.  
The results for the variations between each weight strategy in the experiment and all CFSA tasks are presented in Figure~\ref{fig:RQ4}. The effectiveness indicator is determined by the percentage of scenarios that exceed the trade-off baseline constructed by Fairea. 


Overall, the [0.6-0.4] strategy shows the best results, with CFSA beats 84.25\% of the trade-off baseline cases. 
In real-world deployment, the requirements for fairness and performance may vary depending on task-specific goals. Software engineers can explore different strategies and evaluate their effectiveness to determine the most appropriate strategy for a given task.


\textbf{Ans. to RQ4:} The balance between fairness and performance can be adjusted by the combination strategy that alters the balance. In general, the weighting strategy of [0.6-0.4] can be a starting weight to explore the best weighting strategy based on factors like the number of features and ML algorithm. 

\paragraph{\textbf{RQ5: Is CFSA method efficient in handling multiple sensitive attributes?}}
\label{RQ:RQ5}

The first four RQs examine the bias reduction of CFSA based on a single sensitive attribute, which is the current focus of existing fairness literature~\cite{zhang2022longitudinal, biswas2020machine, biswas2021fair, chakraborty2020fairway, hort2021fairea, chakraborty2021bias, zhang2021fair}. This research question assesses CFSA's effectiveness in dealing with multiple sensitive attributes, which is a common fairness question in real-world~\cite{mehrabi2021survey}. 

\begin{figure}[!htbp]
	\centering
	\includegraphics[width=0.45\textwidth, height=0.35\textwidth]{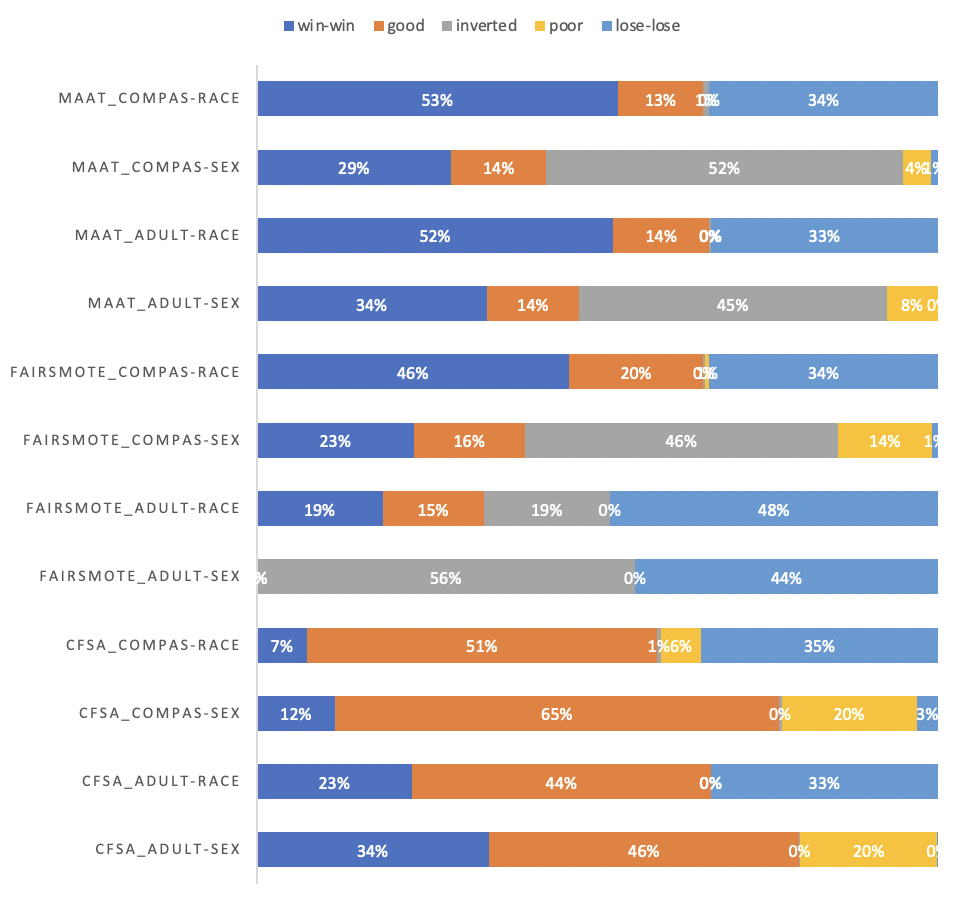}
	\caption{Distribution of CFSA, MAAT, and Fair-SMOTE's effectiveness in multi-attribute tasks. CFSA generally achieves better results (70.5\%) compared to MAAT (55.7\%) and Fair-SMOTE (48.7\%) in terms of good or win-win trade-off in mitigation cases.}
	\label{fig:RQ5}
\end{figure}

We compare CFSA with the MAAT and Fair-SMOTE, the only two approaces that are capable of handling multiple sensitive attributes to the best of our knowledge, in two multi-attribute tasks (i.e., Adult and Compas), still using LR, SVM, and RF as the base model. When training our CFSA, one fair model is built focusing on one sensitive attribute as well as one performance model, which are then averaging strage is used for ensemble learning. With the use of 3 base models, 2 datasets, and 3 methods, and 50 repeated runs, we have a total of 900 mitigation cases ($3\times2\times3\times50=900$). The results, shown in Figure ~\ref{fig:RQ5}, indicate that CFSA had a higher proportion of good or win-win trade-offs compared to other methods and fewer poor or lose-lose trade-offs. For example, CFSA, in the Adult task, outperforms the trade-off baseline for race in 80\% of cases, compared to 33.3\% for Fair-SMOTE and 66.3\% for MAAT.




\textbf{Ans. to RQ5:} CFSA can decrease bias for multiple sensitive attributes. It outperforms state-of-the-art methods by beating the trade-off baseline in 70.5\% of cases, outperforming MAAT and Fair-SMOTE by 14.75\% and 21.75\%, respectively. 

\section{Related Work}
\label{sec:rel-work}
This section presents the two lines of works that are related to this study. 
ML systems are a subset of AI systems, so we first discuss the works related to AI testing.
Then, we discuss the works that are specific to testing and improving fairness in AI systems.

\subsection{AI Testing}

Although demonstrating the potential of tackling important tasks, AI systems still require sufficient amount of effort to ensure their quality from various aspects, including correctness, robustness, security, privacy, etc.
Asyrofi et al.~\cite{crossasrpp} leverage differential testing to synthesize speech inputs using speech-to-text systems to test the correctness of speech recognition systems and show that these inputs can be used to improve the performance of the systems under test~\cite{asrevolve}.
The robustness testing tries to evaluate how AI systems behave when small perturbations are introduced to the inputs. 
Researchers have conducted robustness testing on various AI systems, e.g., computer vision~\cite{sensei,DeepXplore}, code models~\cite{alert,MHM}, reinforcement learning~\cite{rl-attack,wu2021adversarial}, etc.

Researchers have built a wide range of tools to test various AI systems.
Motivated by the usage of code coverage metrics (e.g, line coverage, branch coverage, etc) in testing conventional software systems, 
Pei et al.~\cite{DeepXplore} propose DeepXplore, a tool that uses neuron coverage as a guidance to generate test cases for deep neural networks.
The following researchers extend this work by proposing structural neuron coverage metrics, e.g., neuron boundary coverage, etc. 
These metrics are the foundation for a series of AI testing tools, including DeepHunter~\cite{DeepHunter}, DeepGauge~\cite{DeepGauge}, DeepCT~\cite{DeepCT}, DeepTest~\cite{DeepTest}, etc.
However, recent studies~\cite{li2019structural,harel2020neuron,9376208,FSE_Yan,rl_coverage,9825775} also reveal that neuron coverage may not be effective enough to expose the vulnerabilities of AI systems.
Researchers also explore other metrics to test AI systems.
Gao et al.~\cite{sensei} propose Sensei, a fuzz testing tool that uses genetic algorithms to synthesize inputs to test and improve the robustness of computer vision systems.
Zhang et al.~\cite{deeproad} utilize generative adversarial networks (GANs) to generate driving scenes with various weather conditions to test autonomous driving systems.
We refer readers to~\cite{9000651} for a comprehensive survey of works on AI testing.

\subsection{AI Fairness Testing and Improvement}
A recent survey by Chen et al.~\cite{survey} provides a comprehensive overview of the works on fairness in AI software. 
Beyond the five baselines evaluated in our study, there has been a growing number of works that aim to improve the fairness of AI systems.
Zhang et al.~\cite{10.1145/3377811.3380331} propose a white-box testing technique that leverages adversarial sampling to generate test cases to uncover and repair the fairness violations in DNN-based classifiers.
Zheng et al.~\cite{NeuronFair} design NeuronFair to identify biased neurons and conduct interpretable white-box fairness testing.
Zhang et al.~\cite{10.1145/3460319.3464820} conduct gradient search to improve the efficiency of generating fairness test cases. 
Fan et al.~\cite{10.1145/3510003.3510137} use genetic algorithm to conduct explanation-guided fairness testing. 
Some works focus on improving the natural language processing (NLP) systems.
Asyrofi et al.~\cite{biasfinder} propose BiasFinder, a tool that use metamorphic testing to generate test cases to uncover fairness violations in sentiment analysis.
BiasRV~\cite{biasrv} is based on BiasFinder and can verify fairness violations at runtime. 
Sun et al.~\cite{10.1145/3377811.3380420} uncover bias in machine translation systems. 
Ezekiel et al.~\cite{9678017} use context-free grammar to synthesize inputs to test the fairness of NLP systems.
Researchers also put effort into improving the fairness of AI systems.
Max et al.~\cite{hort2021fairea} design Fairea, a model behaviour mutation approach to benchmark the bias mitigation methods.
Gao et al.~\cite{FairNeuron} model the problem of balancing fairness and accuracy as an adversarial game and propose FairNeuron that can strategically select neurons to improve AI fairness.
Zhang and Sun~\cite{10.1145/3540250.3549103} adaptively improve model fairness based on causality anaysis. 

\section{Conclusion and Future Work}
\label{sec:conclusion}

In this paper, we present CFSA, a method to tackle the root causes of bias in ML software through counterfactual thinking. A thorough evaluation shows that CFSA surpasses existing bias reduction techniques from both the fields of ML and SE significantly by improving fairness while maintaining competitive performance. The successful implementation of CFSA provides possibilities for further exploration into software fairness for fair software engineering development. In the future, we aim to broaden the scope of this work to include text mining and image processing. Additionally, we intend to improve our approach by incorporating new evaluation systems and utilizing industry datasets.

\bibliographystyle{ACM-Reference-Format}
\bibliography{reference}

\end{document}